# Similarity Data Item Set Approach: An Encoded Temporal Data Base Technique

M.S.Danessh, C. Balasubramanian and K. Duraiswamy

*Abstract* - Data mining has been widely recognized as a powerful tool to explore added value from large-scale databases. Finding frequent item sets in databases is a crucial in data mining process of extracting association rules. Many algorithms were developed to find the frequent item sets. This paper presents a summary and a comparative study of the available FP-growth algorithm variations produced for mining frequent item sets showing their capabilities and efficiency in terms of time and memory consumption on association rule mining by taking application of specific information into account. It proposes pattern growth mining paradigm based FP-tree growth algorithm, which employs a tree structure to compress the database. The performance study shows that the anti- FP-growth method is efficient and scalable for mining both long and short frequent patterns and is about an order of magnitude faster than the Apriority algorithm and also faster than some recently reported new frequent-pattern mining.

—————————  ◆  —————————

## 1 INTRODUCTION

One of the currently fastest and most popular algorithms for frequent item set mining is the FP-growth algorithm. It is based on a prefix tree representation of the given database of transactions (called an FP-tree), which can save considerable amounts of memory for storing the transactions. The basic idea of the FP-growth algorithm can be described as a recursive elimination scheme in a preprocessing step delete all items from the transactions that are not frequent individually i.e., do not appear in a user-specified minimum number of transactions. Recourses to process the obtained reduced (also known as projected) database, remembering that the item sets found in the recursion share the deleted item as a prefix. On return, remove the processed item also from the database of all transactions and start over, i.e., process the second frequent item etc. In these processing steps the prefix tree, which is enhanced by links between the branches, is exploited to quickly find the transactions containing a given item and also to remove this item from the transactions after it has been processed[4][7].

The Apriori heuristic achieves good performance gained by (possibly significantly) reducing the size of candidate sets [3]. However, in situations with a large number of frequent patterns, long patterns, or quite low minimum support thresholds, compact data structure, called frequent-pattern tree, or FP-tree in short is constructed, which is an extended prefix-tree structure storing crucial, quantitative information about frequent patterns. To ensure that the tree structure is compact and informative only frequent length-1 items will have nodes in the tree, and the tree nodes are arranged in such a way that more frequently occurring nodes will have better chances of node sharing than less frequently occurring ones. This experiments show that such a tree is compact and it sometimes orders of magnitude smaller than the original database [7]. Subsequent frequent-pattern mining will only need to work on the FP-tree instead of the whole data set. The properties of FP-tree are thoroughly studied [10]. Also, it point out the fact that, although it is often compact, FP-tree may not always be minimal. Some optimizations are proposed to speed up FP-growth which is a technique to handle single path FP-tree has been further developed for performance improvements.

A database projection method has been developed in Section 2 to cope with the situation when an FP-tree cannot be held in main memory the case that may happen in a very large database.

Extensive experimental results have been reported. Thus examine the size of FP-tree as  Well as the turning point of FP-growth on data projection to building FP-tree[9]. The main step is described in Section 3, namely how an FP-tree is projected in order

1. PG student, Dept of CSE, K. S. R. College of Technology,Tiruchengode, Tamilnadu, India.
2. Asst.Professor,Dept of CSE, K.S.R. College of Technology, Tiruchengode, Tamilnadu, India.
3.Dean(academic),Dept of CSE, K.S.R. College of Technology,Tiruchengode, Tamilnadu, India.

to obtain an FP-tree of the (sub) database containing the transactions with a specific item (though with this item removed). It projection step is the most costly in the algorithm and thus it is important to find an efficient way of executing it. The considers how a projected FP-tree may be further pruned using a technique that has been called FP-Growth [4].Such pruning can sometimes shrink the FP-tree considerably and thus lead to much faster projections[8].Finally comparing with my implementations of the FP tree growth and anti-FP tree growth algorithms.

## 2. RELATED RESEARCH WORKS FOR TEMPORAL MINING

The purpose of any data mining process is to identify the important associations among items. An association rule is represented by an activity related to some other activity, for example the purchasing of one product when another product purchased. Several mining algorithms have been proposed to find the association rules from the transactions. The large itemsets were identified to find the association rules. First, the itemsets which satisfy the predefined minimum support were identified and these were called the large itemsets.

Then the association rules were identified from these itemsets [5][6]. The association rules which satisfy the predefined minimum confidence were the association rules produced as the output. The concept of valid time was used to find out the time interval during which a transaction is active. Time interval expansion and mergence was performed which gives importance to the time at which a transaction had taken place. FP-growth proposed in the last section is essentially a main memory-based frequent pattern mining method. However, when the database is large, or when the minimum support threshold is quite low, it is unrealistic to assume that the FP-tree of a database can fit in main memory

A disk-based method should be worked out to ensure that mining is highly scalable. In this section, methods are developed to first partition the database into a set of projected databases, and then for each projected database, construct and mine its corresponding .This takes place before the application of the FP tree growth mining algorithm to identify the temporal association rules [11]. Frequent pattern mining is often regarded as advanced querying where a user specifies the source dataset, the minimum support threshold, and optionally pattern constraints within a given constraint model. A significant amount of research on efficient processing of frequent pattern queries has been done in recent years, focusing mainly on constraint handling and reusing results of previous queries in the context of frequent itemsets and sequential patterns.

```
construct_fptree(database D, flist FList)
input : database D, F-list FList;
output : FP-tree FPtree;
{
1:while not eof(D) do
2: tranline = read_trans(D);
3: begin
4: add all ancestors of each item in tranline
5: removing any duplicates in tranline
6: end
7: o_trans = get_ordered_trans(Flist, tranline);
8: insert_fptree(FPtree, o_trans);
9: end while
}
```
**Pseudo code of FP-tree construction**

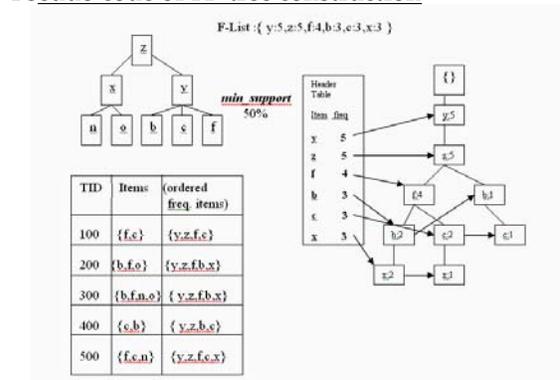

**Fig.1. Example of FP-tree construction**

The FP-tree also has a frequent-item header table that holds the head of the node-links, which connect nodes of same item in FP tree. The node-links facilitate item traversal during the mining of frequent patterns.The two level taxonomy with ancestors depicted on the upper left of the figure, and the transaction database is given below it. The minimum support is set to three (50%). After the first pass, the F-list is determined. Notice that ancestor is also included in the F-list although its descendants are infrequent items. After the transactions are reordered, they are inserted one by one into the FP-tree.

## 3. TREE BASED GENERALIZED ASSOCIATION RULE MINING ALGORITHMS

In purpose algorithm based structure called FP-tree growth but the FP-tree also includes the taxonomy information. The proposed two methods to traverse the FP-tree in order to extract the frequent patterns which will be used to generate the anti-FP tree growth association rules.

## 3.1 The frequent-pattern growth algorithm

The *FP-growth* mining process scans the FP-tree of *DB* once and generates a small pattern-base *Bai* for each frequent item *ai*(ai 's node-links), each consisting of the set of transformed prefix paths[4][6]. Frequent pattern mining is then recursively performed on the small pattern-base Bai (Bai pattern-base node-links) by constructing a conditional FP-tree for Bai. As reasoned in the analysis of Algorithm 1, an FP-tree is usually much smaller than the size of DB. Similarly, since the conditional FP-tree, "FP-tree | ai ", is constructed on the pattern-base Bai , it should be usually much smaller and never bigger than Bai . Moreover, a pattern-base Bai is usually much smaller than its original FP-tree, because it consists of the transformed prefix paths related to only one of the frequent items, ai . Thus, each subsequent mining process works on a set of usually much smaller pattern-bases and conditional FPtrees. Moreover, the mining operations consist of mainly prefix count adjustment, counting local frequent items, and pattern fragment concatenation. This is much less costly than generation and test of a very large number of candidate patterns. Thus the algorithm is efficient.

**Method:** call FP-growth(FP-tree, null).
**Procedure FP-growth**(Tree, $\alpha$)
{(1) if Tree contains a single prefix path // Mining single prefix-path FP-tree
(2) then {
 (3) let P be the single prefix-path part of Tree;
(4) let Q be the multipath part with the top branching node replaced by a null root;
(5) for each combination (denoted as $\beta$) of the nodes in the path P do
(6) generate pattern $\beta \wedge \alpha$ with support = minimum support of nodes in $\beta$;
(7) let freq pattern set(P) be the set of patterns so generated; }
(8) else let Q be Tree;
(9) for each item ai in Q do { // Mining multipath FP-tree
(10) generate pattern $\beta$ = ai U $\alpha$ with support = ai .support;
(11) construct $\beta$'s conditional pattern-base and then $\beta$'s conditional FP-tree Tree$\beta$ ;
(12) if Tree $\beta$ = 0
 (13) then call FP-growth(Tree$\beta$, $\beta$);
(14) let freq pattern set(Q) be the set of patterns so generated; }
(15) return(freq pattern set(P) U freq pattern set(Q) U (freq pattern set(P)×freq pattern set(Q)))
}

## 3.2 The frequent-pattern tax algorithm

FPtax traverses nodes in the FP-tree starting from the least frequent item in F-list. While visiting each node, BU-FPtax also collects the prefix-path of the node, which is the set of items on the path from the suffix node to the root of the tree.FPtax also stores the count on the node as the count of the prefix path pattens[2]. The conditional pattern base is a small database of frequent patterns that co-occur with the item.

Then BU-FPtax creates small FP-tree from the conditional pattern base called conditional FPtree. During each iteration, a new frequent itemset is generated by adding the suffix to the itemset from the previous iteration. BUFPtax also maintains a list of ancestors anclist for the items in the current itemset.

procedure BU-FPtax(FPtree, X, anclist);
input : FP-tree Tree, itemset X, itemset anclist;
{
1:for each item y (bottom-up order)
in the header of FPtree do
2: if(y is in anclist) then continue;
//Filtering 1
3: if(ancestors of y is in Y) then continue;
//Filtering 2
4: generate pattern Y = y U X with
support = y.support;
5: begin
6: add ancestors of y to anclist;
7: end
8: cond_pbase = construct_cond_pbase(Tree,y);
9: Y-Flist = sort_cond_pbase(cond_pbase);
10: Y-Tree = construct_fptree(cond_pbase,Y-FList);
11: if (Y-Tree is not NULL) then
12: BU-FPtax(Y-Tree, Y, anclist);
13: end if
14: end for
}
  Pseudo code for Bottom-Up FP-tax(BU FP tax)

Here it proposes two kind of filtering methods utilizing the characteristics of ancestor-descendant relation to reduce the search space. The foundation for the optimizations is given in the cumulate paper [8], which states the following observations: Do not need to count any itemset which contains both an item and its ancestor because the support is contained in the ancestor's support [9].

Pruning of such itemset is sufficient to ensure that it never generate itemsets in subsequent iterations which contain both an item and its ancestor. Two kinds of filtering methods are needed to remove itemsets that contain both a completely item and its ancestors. Pruning itemset whose item is already in the anclist, the anclist contains all ancestors examined so far for the current suffix.

Thus it is sufficient to check the membership of the anclist against the item to make sure that no ancestors of the item in the itemset. Pruning items whose ancestors is already included in the Itemset The second filtering is needed in order to prune the descendants of the items in the itemset. The first filtering only checks the ancestors of the item. It is also obvious that a direct descendant of items in the itemset is not needed. When BU-FPtax encounters such conditions, the iteration is stopped and BU-FPtax processes the next item. The process is recursively iterated until no conditional pattern base can be generated and all frequent patterns that contain the item are discovered.

### 3.3 TOPDOWN TRAVERSAL METHOD

In top-down algorithm Top-Down FP-tax is inspired by algorithm Top-Down FP-growth. The advantage of this method is no need to construct the conditional pattern bases and the sub trees. The filtering methods similar with Bottom-Up FP-tax are also employed to avoid the unnecessary traversals of items which occur with their ancestors. The recursive function TD-FPtax is TD FPtax eliminates the generation of conditional pattern bases by directly reconnecting the node links of the FP-tree. During the reconnection, TD-FPtax also modifies the count information in the FP-tree nodes. For each new suffix extension a new header able is created. The header table contains items prior to the item in the previous header table[11]. Then following the node-links of the prefix paths to the root node are traversed. During the traversal, the node links of the FP-tree nodes are reconnected to header table. The counts in header table and in the FP-tree node are also modified relative to its co-occurrence with the process have recursively iterated[3].

### 4. PERFORMANCE EVALUATION

**The Encoding Method**

The presentation of database is an important consideration in almost all algorithms. The most commonly used layout is the horizontal database layout and vertical one. In both layouts, the size of the database is very large. A large database to be transformed into a smaller one with all properties of its original layout is expected. Database encoding is a new presentation, which can reduce the size of database and improve the efficiency of algorithms. Instead of maintaining a large table in the transaction database, one table is created with only two columns [9].

The first one is the transaction identifier and another is for the entire items that occur in the transaction. All items in one transaction are converted into only one number that has all properties of these items. By this way, the new database is much smaller than the previous one and can be loaded into memory easily. So the cost of memory is reduced [12][13].According to the assumption that only one number represents an itemset, when converting an itemset into a number, a measure attribute is defined, which is a numerical attribute associated with every item in each transaction in the database layout. A binary number expresses a numerical attribute, that is, those items that are occurring in one transaction are depicted with 1 and all the other items are represented with 0. The transaction measure value, denoted as tmv ($I_p$, $T_q$), is a value of a measure attribute related to an item $I_P$ in a transaction $T_q$. tmv ($I_P$, $T_q$)= 0 means item $I_P$ does not occur in the transaction $T_q$, while tmv ($I_P$, $T_q$)= 1 means item $I_P$ occurs in the transaction $T_q$. In table 1, for example, tmv ($I_4$, $T_1$) is equal to 1. Any item $I_P$ in the set of items is encoded as one prime number, denoted as E ($I_P$). Prime numbers are used because any number except 1 and themselves cannot divide them. For any item $I_P$ in the transaction $T_q$, a new measure denoted as M ($I_P$, $T_q$) is equal to the product of tmv ($I_P$, $T_q$) and its encoding number E ($I_P$) is assigned. This value is gotten by Eq. (1). After this step, for all $I_P$ and $T_q$, if M ($I_P$, $T_q$) equal to 0, then convert M ($I_P$, $T_q$) into 1. This operation is described in Eq. (2). For any transactions, the value $M_{T_q}$ is equal to the multiplication of all M ($I_P$, $T_q$). The value of $M_{T_q}$ is represented in Eq. (3).

$$M (I_P, T_q) = tmv(I_P, T_q) \times E(I_P) \qquad (1)$$

$$\text{For all}(I_P, T_q) \text{ If } M (I_P, T_q) = 0 \Rightarrow M(I_P, T_q) = 1 \qquad (2)$$

$$M_{T_q} = \Pi (I_P, T_q) \qquad (3)$$

For any itemset I=($I_{p1}$, $I_{p2}$, …, $I_{pn}$), there is one value denoted as $M_I$ is equal to the multiplication of all E($I_P$) if its $I_P$ occur in I, as described in Eq. (4). The value $M_I$ shows the number corresponding to itemset I. And then this number can be used instead of itemset I.

$$M_I = \Pi \, E (I_P) \qquad (4)$$
$$\forall \, I_P \in I$$

With this encoding, instead of maintaining all tmv ($I_P, T_q$) for every item and transaction, the value $M_I$ can be stored for every transaction.

### 5. ANTI FP TREE GROWTH ALGORITHM

The Anti FP-growth mining process scans the FP-tree of *DB* and generates a small pattern-base $B_{ai}$ for each frequent item *ai* , each consisting of the set of transformed prefix paths of ai .Frequent pattern mining is then recursively performed on the small pattern-base Bai by constructing a conditional FP-tree for Bai. As reasoned in the analysis of Algorithm, FP-tree is usually much smaller the size of the Database. Moreover, the encode database techniques have mining operations consist of mainly prefix count adjustment, counting local frequent items, and pattern fragment concatenation. This is much less costly than generation and test of a very large number of candidate patterns. Thus the algorithm is efficient.

The first example is on mining frequent closed itemsets. Since the frequent pattern mining often generates a very large number of frequent itemsets, it hinders the effectiveness of mining since users have to sift through a large number of mined rules to find useful ones. An mine frequent closed itemsets, where an itemset $\alpha$ is a closed itemset if there exists no proper superset of $\alpha$ that has the same support as $\alpha$ in the database. Mining frequent closed itemsets has the same power as mining the complete set of frequent itemsets, but it may substantially reduce redundant rules to be generated and increase the effectiveness of mining.

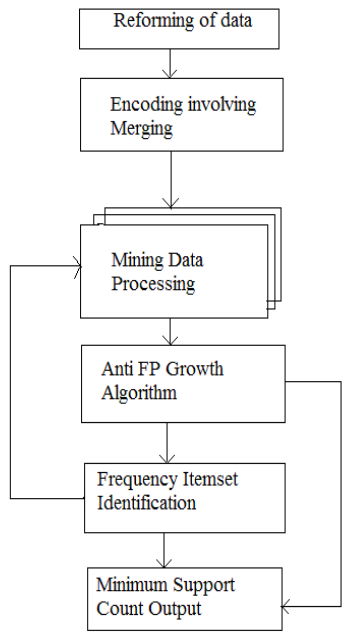

Fig.1. The proposed classification based Encoded temporal mining

The proposed methodology for classification based temporal mining has the following steps.

1. **Data Reforming:** It describes the way in which the time intervals are set up. The choice of the time depends upon the user and the application.
2. **Encoding Involving merging:** This step encodes the temporal database for the given time in such a way that redundancy is avoided by merging.
3. **Mining data processing:** This identifies the large item sets by applying the priority based temporal mining method.
4. **Anti-FP growth Algorithm:** It identifies the association rules in each interval and applies transitive property to find out the relationship between the associations rules resulting from different time intervals. It also expands the time intervals, if the time intervals are continuous and hold the same association rules which satisfy the minimum confidence. It identifies the probability for the classes involved which leads to temporal rules for easy solutions.

**Method:** Anti-FP-growth (FP-tree, null).

**Procedure Anti FP-growth**(Tree, $\alpha$)

{(1) if Tree contains a single prefix path // Mining single prefix-path FP-tree
(2) then
{
(3) let P be the single prefix-path part of Tree;
(4) let Q be the multipath part with the top branching node replaced by a null root;
(5) for each combination (denoted as β) of the nodes in the path P do
{
(6) generate pattern β ^ $\alpha$ with support = minimum support of nodes in β;
(7) let freq pattern set(P) be the set of patterns so generated;
}
(8) else let Q be Tree;
(9) for each item ai in Q do { // Mining multipath FP-tree
(10) generate pattern β = ai U $\alpha$ with support = ai .support;
(11) construct β's conditional pattern-base and then β's conditional FP-tree Treeβ ;
(12) if Tree β = 0
 (13) then call FP-growth(Treeβ, β);
(14) let freq pattern set(Q) be the set of patterns so generated; }
(15) return(freq pattern set(P) U freq pattern set(Q) U (freq pattern set(P)×freq pattern set(Q)))
(16) else if Tree β ≠ 0
 (17) then call FP-growth(Treeβ, β);
(18) let freq pattern set(Q) be the set of patterns haven't generated but minimum support count on between the intervals (Treeβ, β);
} (19) return(freq pattern set(P) ∩ freq pattern set(Q) ∩ (freq pattern set(P)×freq pattern set(Q))) }

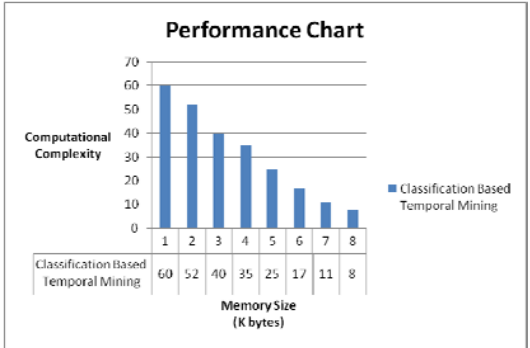

Fig.2.Performance of classification Encoded based mining

The classification based temporal mining methodology has better performance in terms of the logic used. The performance of this method increases with increase in the number of tuples within a given time interval. The computational complexity decreases with increase in the number of transactions. This is depicted in fig.2 as follows.Fig.3 shows the usage of memory before and after encoding. It is also found that the encoding method leads to faster generation of temporal association rules. The algorithm involved in mining association rules with frequent items is efficient for large databases as shown in figure 4.

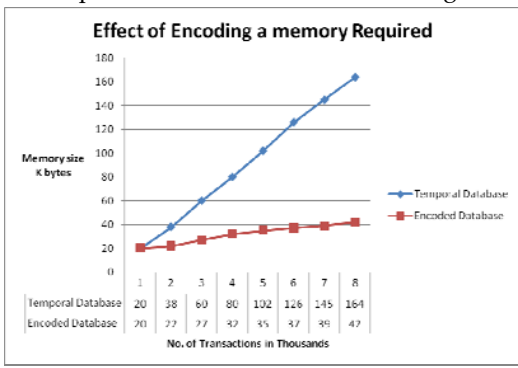

Fig. 3.Memory usage graph

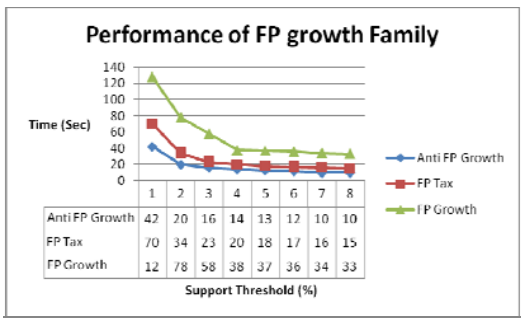

Fig.4.Performance of Anti FP Growth algorithms

| Algorithms | Data set's approaches | Surveys(Memory) |
|---|---|---|
| FP Tree Growth | Top Down approaches | 55% (Temporal data) |
| FP-tax Tree Growth | TD & BU approaches | 35% (Spatio- Temporal data) |
| Anti-FP Tree Growth | Bottom Up approaches | 20% (Encoded Temporal data) |

Table 1.Surveys of Purposed Algorithms

## 6. CONCLUSION AND FUTURE ENHANCEMENTS

This paper projected the impact of the FP growth family of algorithms on an encoded database for association rule mining. Each of the algorithms has a different impact and produces effective results. The effect of FP growth and anti-FP growth algorithms on a temporal data base which has been encoded by an encoding method has more openings to provide advantageous results in terms of lower complexities of time and space. Also, investigating the problem of quantitative association rules with weighted items may be taken as the subject for research in the future.

**Acknowledgement**:
Procs.No.2006/459/111/Ph.D./RDR/SLM, the author would like to express that above process is going on successfully.

# BIBLIOGRAPHY

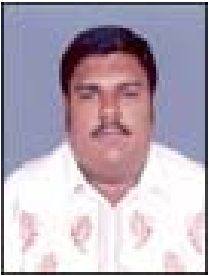

**DANESSH MOORTHY,** Student of Anna University Coimbatore perceiving ME degree in the Department of Computer Science and Engineering. He completed his BE degree in the Department of Computer Science and Engineering from Anna University Chennai. His area of interest includes DataMining, DataBase, Mobile Computing and Computer Networks.

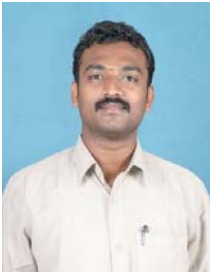

**C. BALASUBRAMANIAN,** Research scholar of Anna University, Chennai, is working as an Asst.Professor at K.S.Rangasamy college of technology, Tiruchengode, Tamilnadu. He completed his Bachelor's degree in Engineering from Madurai Kamaraj University, Tamilnadu and Masters Degree in Engineering from Annamalai University, Tamilnadu. His research interests include Database Management Systems, Temporal Mining and Computer Networks**.**

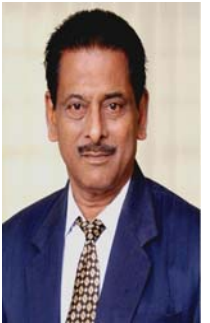

**DR. K. DURAISWAMY** received his PhD degree in computer science and Engineering from Anna University, Chennai. He is now working as a Dean (Academic) at K.S.Rangasamy College of Technology, Tiruchengode and Research guide at Periyar University, Salem, Anna University, Chennai and Coimbatore. His research interest includes image processing, Mobile Computing, Database Management Systems, Data warehousing, Digital Signal processing, Artificial Intelligence and Neural Networks. He has published more than 250 technical papers at various National International conferences and Journals.